\documentclass{JHEP3}
\usepackage{amsmath,amssymb}
\usepackage{multirow}
\usepackage{bm}
\usepackage{epsfig}
\usepackage{graphicx}
\usepackage{amsmath}
\usepackage{yfonts}
\usepackage{subfigure}

\makeatletter \@addtoreset{equation}{section}

\newcommand{\be}{\begin{equation}}
\newcommand{\ee}{\end{equation}}
\newcommand{\ba}{\begin{eqnarray}}
\newcommand{\ea}{\end{eqnarray}}

\newcommand{\bsubeq}{\begin{subequations}}
\newcommand{\esubeq}{\end{subequations}}

\def\lsim{\mathrel{\rlap{\lower3pt\hbox{\hskip1pt$\sim$}}
     \raise1pt\hbox{$<$}}} 
\def\gsim{\mathrel{\rlap{\lower3pt\hbox{\hskip1pt$\sim$}}
     \raise1pt\hbox{$>$}}}
\def\HLS1{HLS$_1$}
\newcommand{\textoverline}[1]{$\overline{\mbox{#1}}$}

\DeclareFontFamily{U}{rsf}{}
\DeclareFontShape{U}{rsf}{m}{n}{
  <5> <6> rsfs5 <7> <8> <9> rsfs7 <10-> rsfs10}{}
\DeclareMathAlphabet\Scr{U}{rsf}{m}{n}

\allowdisplaybreaks

\title{Holographic equations of state and astrophysical compact objects}

\author{Youngman Kim
        \\
Asia Pacific Center
for Theoretical Physics and Department of Physics, Pohang University
of Science and Technology, Pohang, Gyeongbuk 790-784, Korea
\\  E-mail: \email{ykim@apctp.org}}

\author{Chang-Hwan Lee
\\
Department of Physics, Pusan National University, Busan 609-735, Korea
\\ E-mail: \email{clee@pusan.ac.kr}}

\author{Ik Jae Shin, Mew-Bing Wan
\\
Asia Pacific Center
for Theoretical Physics, Pohang, Gyeongbuk 790-784, Korea
\\ E-mail: \email{geniean@apctp.org, mbwan@apctp.org}}

\abstract{ We solve the Tolman-Oppenheimer-Volkoff equation using an equation of state (EoS) calculated
in holographic QCD. The aim is to use compact astrophysical objects like neutron stars as an indicator
to test holographic equations of state. We first try an EoS from a dense D4/D8/\textoverline {D8} model.
In this case, however, we could not find a stable compact star,
 a star satisfying pressure-zero condition with a radius $R$, $p(R)=0$,
within a reasonable value of the radius. This means that the EoS from the D4/D8/\textoverline {D8} model
may not support any stable compact stars or may support one whose radius is very large.
This might be due to a deficit of attractive force from a scalar field or two-pion exchange
in the D4/D8/\textoverline {D8} model. Then, we consider  D4/D6 type models
with different number of quark flavors, $N_f=1,2,3$.
Though the mass and radius of a holographic star is larger than those of normal neutron stars, the D4/D6 type EoS renders a stable compact star.}

\keywords{Gauge/gravity duality, dense matter, compact stars}

\begin{document}

\section{Introduction}

The approaches based on the Anti de Sitter/conformal field theory (AdS/CFT) correspondence \cite{Maldacena:1997re,Gubser:1998bc,Witten:1998qj} are showing many interesting possibilities to explore strongly interacting systems such as dense baryonic matter, stable/unstable nuclei, and strongly interacting quark gluon plasma.
The basic strategy is to introduce an extra space, which roughly corresponds to the
energy scale of  4D boundary field theory, and try to construct a 5D
holographic dual model that captures certain non-perturbative aspects
of strongly coupled field theory.
There are in general two different
routes to modeling holographic dual of quantum chromodynamics (QCD). One
way is a top-down approach based on stringy D-brane configurations. The other way is so-called a bottom-up approach, in which a 5D
holographic dual is constructed from a boundary field theory such as QCD. For reviews on holographic QCD we refer to \cite{review_hQCD}.


Neutron stars are the most important astrophysical compact objects in which the dense matter physics can be explored and tested.
Since one cannot obtain the information on the inner structure of neutron star directly from observation, mainly masses and radii of neutron stars are being considered. Due to the uncertainties in dense matter physics at high densities, various equations of states have been suggested \cite{Lattimer:2006xb}. One of the main problems lies in the fact that most of the microscopic theories are based on the data at low densities, only up to normal nuclear matter density. Hence, in order to guide high density behavior, new approaches which are based on QCD fundamental symmetries have been suggested. This work is along this line of work even though it is still far from realistic.

Most accurate measurements on the neutron star masses has been done in double neutron star binaries and well measured neutron star masses in these systems are consistent with $<1.5 M_\odot$ \cite{Lattimer:2006xb}. However, recently, a $1.97\pm 0.04 M_\odot$ neutron star has been observed in a neutron star - white dwarf binary J1614$-$2230 using Shapiro delay \cite{Demorest:2010bx}. In addition, both masses and radii of neutron stars in low mass X-ray binaries have been estimated and more data will come soon \cite{Guver:2011qw,Guver:2011js}.
These observations already ruled out some sets of equations of state. In the future, once the advanced LIGO is operating, gravitational waves from neutron star binary inspirals are expected to be detected. These observations may be able to provide clues on the inner structure of neutron stars.


In this work, to see what the equation of state (EoS) from holography says about nature,
 we solve the Tolman-Oppenheimer-Volkoff (TOV) equation using the EoS derived
from a  top-down model. We use compact astrophysical objects like neutron stars as a testing agent.
We first take an EoS from a cold and dense D4/D8/\textoverline {D8} model.
With this EoS, we cannot find any stable compact stars with a reasonable radius.
This might be due to a weakness of attractive force in the D4/D8/\textoverline {D8}~\cite{Sakai:2004cn}.
In nuclear physics, the long range attractive force is mostly given by one-pion exchange, while the intermediate
range $\sim 1$ fm attraction is dominated by a scalar field or two-pion exchange.
Note that in the D4/D8/\textoverline {D8}~\cite{Sakai:2004cn} a scalar that couples to baryon fields is missing.
Then, we consider a D4/D6 type model
with different number of quark flavors, $N_f=1,2,3$.  Since the D4/D6 type models have both vector and scalar bulk fields,
we might expect that the EoS calculated in this model could support a stable compact star.
We find that the D4/D6 type EoS allows a stable
 compact star, though the mass and radius is a bit larger than observed neutron stars.

One cautionary remark is that a simple classical
  analysis in the holographic dual model, both top-down and bottom-up,
captures only the leading $N_c$ contributions, and we are bound to suffer from sub-leading
corrections. For instance for low densities we may neglect a back-reaction from a bulk U(1) gauge field
whose boundary value defines a chemical potential and corresponding density. However, at high densities, though we
have no clear-cut number to tell low density from high density, the back-reaction will not be negligible.

\section{D4/D6 model}
 We briefly summarize the D4/D6 system \cite{Kruczenski:2003uq} and comment on its extended version in dense matter with $N_f\ge 1$ \cite{Seo:2008qc, Kim:2009ey}.
 The model contains $N_c$ number of D4 branes and $N_f$ flavor D6 branes whose configurations are given in Table \ref{braneprofile1}.
 \begin{table}[h]
  \begin{center}
  \begin{tabular}{|c|c|c|c|c|c|c|c|c|c|c|}
	\hline
	&t&1&2&3& ($\tau$) & $z$ & $\psi_1$ &$\psi_2$&r&$\phi$\\
	\hline
	D4&$\bullet$&$\bullet$&$\bullet$&$\bullet$&$\bullet$&&&&& \\
	\hline
	D6&$\bullet$&$\bullet$&$\bullet$&$\bullet$&&$\bullet$&$\bullet$&$\bullet$&& \\
	\hline
  \end{tabular}
  \caption{The brane configurations : the background D4 and the probe D6 \label{braneprofile1}}
  \end{center}
 \end{table}
 In the probe limit, the $N_c$ D4 branes are replaced by corresponding type IIA SUGRA background, while $N_f$ D6 branes are treated as probes.
 The solution of confining D4 brane reads
 \begin{eqnarray} \label{backgroundD6}
	ds^2&\!=\!&\Big(\frac{U}{L}\Big)^{3/2}\big(\eta_{\mu\nu}dx^\mu dx^\nu+f(U)d\tau^2\big)+\Big(\frac{L}{U}\Big)^{3/2}\Big(\frac{dU^2}{f(U)}+U^2d\Omega_4^2\Big) \nonumber \\
	&\!=\!&\Big(\frac{U}{L}\Big)^{3/2}\big(\eta_{\mu\nu}dx^\mu dx^\nu+f(U)d\tau^2\big)+\Big(\frac{L}{U}\Big)^{3/2}\Big(\frac{U}{\xi}\Big)^2(d\xi^2+\xi^2d\Omega_4^2)\,, \\
	&&~~~~~~~~~~e^\phi=g_s\Big(\frac{U}{L}\Big)^{3/4}~~~,~~~~F_4=dC_3=\frac{2\pi N_c\epsilon_4}{\Omega_4}\,,
 \end{eqnarray}
 where $f(U)=1-(U_\textrm{KK}/U)^3\,$.
 The D4 branes are compactified on the $\tau$-direction, and $U$ is considered as a radial coordinate of the transverse to the world-volume of branes.
 $\epsilon_4$ and $\Omega_4$ denote the volume form and volume of the unit $S^4$, respectively.
 For convenience, we introduce the dimensionless radial coordinate $\xi$  defined as
 \begin{equation}
	\Big(\frac{U}{U_\textrm{KK}}\Big)^{3/2}\equiv\frac{1}{2}\big(\xi^{3/2}+\frac{1}{\xi^{3/2}}\big)\,.
 \end{equation}
 The parameters in string theory are related to those of the gauge side as
 \begin{equation}
	L^3=\frac{\lambda l_s^2}{2M_\textrm{KK}}~~,~~~~U_\textrm{KK}=\frac{2\lambda M_\textrm{KK}l_s^2}{9}~~,~~~~g_s=\frac{\lambda}{2\pi N_c M_\textrm{KK} l_s}~,
 \end{equation}
 with the 't$\,$Hooft coupling $\lambda\equiv g_\textrm{YM}^2 N_c$.
 To describe the D6 brane configuration, the background metric (\ref{backgroundD6}) is rewritten as
 \begin{equation}
	ds^2=\Big(\frac{U}{L}\Big)^{3/2}\big(\eta_{\mu\nu}dx^\mu dx^\nu+f(U)d\tau^2\big)+\Big(\frac{L}{U}\Big)^{3/2}\Big(\frac{U}{\xi}\Big)^2\big(dz^2+z^2d\Omega^2_2+dr^2+r^2d\phi^2\big)
 \end{equation}
 where $\xi^2=z^2+r^2$.
 The location of D6 brane is described by $r(z)$ with $\phi=0$ and $\tau=\tau_0$.
 The distance between D4 and D6 branes is interpreted as a quark mass; $r(\infty)\sim m_q\,$.
 Then the induced metric on the D6 brane is
 \begin{equation} \label{inducedD6}
	ds_\textrm{D6}^2=\Big(\frac{U}{L}\Big)^{3/2}\eta_{\mu\nu}dx^\mu dx^\nu+\Big(\frac{L}{U}\Big)^{3/2}\Big(\frac{U}{\xi}\Big)^2\Big[(1+{r^\prime}^2)dz^2+z^2d\Omega_2^2\Big]\,,
 \end{equation}
 where the prime denotes the derivative with respect to $z$.
 The embedding configuration of probe brane is governed by Dirac-Born-Infeld (DBI) action.
 The DBI action for D6 brane, in the case of $N_f=1$, becomes
 \begin{eqnarray} \label{DBIactionD6}
	S_\textrm{D6}&\!=\!&-\mu_6\!\int\!d^7\sigma~e^{-\phi}\!\sqrt{-\textrm{det}(\textrm{P}[g]+2\pi\alpha^\prime F)} \nonumber \\
	&\!=\!&-\tau_6\int dt dz\;z^2(1+1/\xi^3)^{4/3}\sqrt{(1+1/\xi^3)^{4/3}(1+{r^\prime}^2)-\tilde{F}^2}
 \end{eqnarray}
 where $\mu_6^{-1}=(2\pi)^6 l_s^7\;,~\tau_6=\mu_6 g_s^{-1}\Omega_2 V_3(U_\textrm{KK}^3/4)$ and $\tilde{F}=(2^{2/3}/U_\textrm{KK})2\pi\alpha^\prime F_{zt}=\tilde{A}_t^\prime\,$.
 Here, for later convenience, more general DBI action including a gauge field $A_t$ is considered.
 The source of the gauge field is the end points of fundamental strings.
 We define the dimensionless quantity $\tilde{Q}$,
 \begin{equation}
	 \frac{\partial\mathcal{L}_\textrm{D6}}{\partial\tilde{F}}=\frac{z^2(1+1/\xi^3)^{4/3}\tilde{F}}{\sqrt{(1+1/\xi^3)^{4/3}(1+{r^\prime}^2)-\tilde{F}^2}}\equiv\tilde{Q}\,,
 \end{equation}
 and $\tilde{Q}$ is related to the number of fundamental strings $Q$ by $\tilde{Q}=Q/(2\pi\alpha^\prime\tau_6)(U_\textrm{KK}/2^{2/3})\,$.
 Here $Q$ is identified with the total quark number of a system.
 The Hamiltonian of the D6 brane is obtained through the Legendre transformation.
 \begin{equation} \label{HamiltonianD6}
  	H_\textrm{D6}=\tau_6\int dz\;\sqrt{(1+1/\xi^3)^{4/3}\left[z^4(1+1/\xi^3)^{8/3}+\tilde{Q}^2\right](1+{r^\prime}^2)}
 \end{equation}

 This D4/D6 model is extended to study dense matter in confined phase \cite{Seo:2008qc} and to investigate a model for transition from nuclear matter to strange matter \cite{Kim:2009ey}.
 Here we briefly review the D4/D6 model in dense matter and refer to \cite{Seo:2008qc, Kim:2009ey, Witten:1998xy} for details.
 In holographic QCD, a compact D4 brane wrapping on the 4-sphere $S^4$ transverse to $\mathbb{R}^{1,3}$ is introduced as a baryon \cite{Witten:1998xy}.
 To describe the compact D4 brane, the $S^4$ part of the background metric (\ref{backgroundD6}) is rewritten as $d\Omega_4^2=d\theta^2+\sin^2\!\theta d\Omega_3^2\,$.
 \begin{table}[h]
  \begin{center}
  \begin{tabular}{|c|c|c|c|c|c|c|c|c|c|c|}
	\hline
	& t & 1 & 2 & 3 & ($\tau$) & $\xi$ & $\theta$ & $\varphi_1$ & $\varphi_2$ & $\varphi_3$ \\
	\hline
	D4 & $\bullet$ & & & & & & $\bullet$ &$\bullet$ & $\bullet$ & $\bullet$ \\
	\hline
  \end{tabular}
  \caption{The brane configuration : the compact D4 \label{braneprofile2}}
  \end{center}
 \end{table}
 The world-volume coordinate for the compact D4 is $(\,t, \theta, \varphi_\alpha)$ where $\theta$ is the polar angle from the south pole.
 To study dense matter, we turn on the time component of the gauge field $A_t$.
 The $SO(4)$ symmetry gives the ansatz that all fields depend on $\theta$, that is, $\xi(\theta)$ and $A_t(\theta)$.
 The induced metric on this D4 brane is
 \begin{equation} \label{inducedD4}
	ds_\textrm{D4}^2=-\Big(\frac{U}{L}\Big)^{3/2}dt^2+(L^3 U)^{1/2}\big[(1+\frac{\dot{\xi}^2}{\xi^2})d\theta^2+\sin^2\!\theta d\Omega_3^2\big]
 \end{equation}
 where $\dot{\xi}=\partial\xi/\partial\theta$.
 The DBI action of the compact D4 brane is
 \begin{eqnarray} \label{DBIactionD4}
	S_\textrm{D4}&\!=\!&-\mu_4\!\int\!d^5\sigma~e^{-\phi}\!\sqrt{-\textrm{det}(\textrm{P}[g]+2\pi\alpha^\prime F)} \nonumber \\
	&\!=\!&-\tau_4\!\int\!dt d\theta\,\sin^3\!\theta\sqrt{(1+1/\xi^3)^{4/3}(\xi^2+\dot{\xi}^2)-\tilde{F}^2}
 \end{eqnarray}
 where $\mu_4^{-1}=(2\pi)^4 l_s^5\;,~\tau_4=\mu_4 g_s^{-1}\Omega_3 L^3(U_\textrm{KK}/2^{2/3})$ and $\tilde{F}=(2^{2/3}/U_\textrm{KK})2\pi\alpha^\prime F_{\theta t}=\dot{\tilde{A}}_t\,$.
 The dimensionless displacement is defined as
 \begin{equation}
	\frac{\partial\mathcal{L}_\textrm{D4}}{\partial \tilde{F}}=\frac{\sin^3\!\theta\,\tilde{F}}{\sqrt{(1+1/\xi^{3})^{4/3}(\xi^2+\dot{\xi}^2)-\tilde{F}^2}}\equiv\tilde{D}\,,
 \end{equation}
  which is given by Chern-Simons (CS) term.
 \begin{equation} \label{CStermD4}
	S_\textrm{D4}^\textrm{(CS)}=\mu_4\!\int\!A_{1}\wedge G_{4}=\tau_4\!\int\!dt d\theta~3\sin^3\!\theta\,\tilde{A}_t
 \end{equation}
 From the equation of motion of the gauge field, $\tilde{D}$ is obtained explicitly.
 \begin{equation}
	\tilde{D}(\theta)=2-(3\cos\theta-\cos^3\theta)
 \end{equation}
 The Hamiltonian of the compact D4 brane is also given by the Legendre transformation.
 \begin{equation} \label{HamiltonianD4}
  	H_\textrm{D4}=\tau_4\int d\theta\;\sqrt{(1+1/\xi^3)^{4/3}(\xi^2+\dot{\xi}^2)}\sqrt{\tilde{D}^2+\sin^6\theta}
 \end{equation}

 We consider the system in which the number of fundamental strings is $Q$.
 Since $N_c$ fundamental strings are attached on each compact D4 brane, there exist $N_B(=Q/N_c)$ number of D4 branes.
 The other end of the fundamental string is located on the D6 brane as a source of the gauge field.
 Since the tension of the string is always larger than that of the brane, the fundamental strings pull both D4 and D6 branes.
 Eventually the length of the string vanishes and the compact D4 and D6 branes meet at a point.
 To be a stable system, the force at the cusp should be balanced: $(Q/N_c)f_\textrm{D4}=f_\textrm{D6}[Q]\,$.
 Force due to the brane tension at the point is given by the variation of its Hamiltonian with respect to the position of contact point.
 \begin{equation}
	f\equiv\frac{\partial H}{\partial\xi_c}\Big|_\textrm{fixed other values}
 \end{equation}

 At zero temperature, energy density $\epsilon$ and pressure $p$ are calculated as
 \begin{equation}
	 \epsilon=\frac{1}{V_3}\Big(\frac{Q}{N_c}\bar{\Omega}_\textrm{D4}+\bar{\Omega}_\textrm{D6}^\textrm{reg.}[Q]\Big)~~~\textrm{and}~~~~~p=\frac{1}{V_3}\Omega_\textrm{D6}^\textrm{reg.}[Q]
 \end{equation}
 where $\Omega$ and $\bar{\Omega}$ denote the grand canonical potential and its Legendre transformed one, respectively.
 These quantities can be obtained as on-shell action and Hamiltonian.
 \begin{equation}
	\Omega\equiv S|_\textrm{on-shell}~~~~,~~~~~~~~\bar\Omega\equiv H |_\textrm{on-shell}
 \end{equation}
 When we calculate the on-shell values of probe D6 branes, the integration result with the infinite range of $z$ diverges.
 Therefore, we should subtract out the zero-density contribution.
 \begin{equation}
	\Omega^\textrm{reg.}[Q]\equiv\Omega[Q]-\Omega[Q=0]
 \end{equation}
 The baryon number density is also given by $\rho=Q/(N_c V_3)\,$.

 If we want to extend the single flavor description to a multi-flavor case $(N_f>1)$, it is convenient to introduce the quantity $q\equiv Q/N_f\,$.
 In this work we consider
 all $N_f$ D6 probe branes have identical asymptotic heights, {\it i.e.}, $N_f$ quark flavors with the same masses.
 In this case we can trivially extend the single flavor case to the multi-flavor one.
 For a fixed $Q$, $Q/N_f$ number of fundamental strings are attached on each D6 probe brane.
  Then the force balancing condition is modified as $(Q/N_c)f_\textrm{D4}=N_f f_\textrm{D6}[q]\,$. The energy density and the pressure are
 \begin{equation} \label{EandPinD4D6}
	 \epsilon=\frac{1}{V_3}\Big(\frac{Q}{N_c}\bar{\Omega}_\textrm{D4}+N_f\,
\bar{\Omega}_\textrm{D6}^\textrm{reg.}[q]\Big)~~~~,~~~~~~~~p=\frac{N_f}{V_3}\,\Omega_\textrm{D6}^\textrm{reg.}[q]\;.
 \end{equation}

\section{Numerical Results}
Now, we use the EoS obtained in holographic QCD models to calculate the mass-radius relation of a compact
star. To this end,
we solve the general-relativistic TOV equation using the holographic EoS:
\begin{eqnarray}
\label{eq:tov}
\frac{dp}{dr} = -\frac{1}{2}(\epsilon+p)\frac{2m+8\pi r^3 p}{r(r-2m)},
\end{eqnarray}
where $r$ is the circumferential radius, and geometric units, $c=G=1$ are used.
The energy density, $\epsilon$, is given by $\epsilon \equiv (E/A+m_b)\rho$, where $E/A$ is the energy per baryon, $m_b=938$ MeV is the mass of the baryon, and $\rho$ is the baryonic number density.
The integrated mass of the compact star is given by $m(r) \equiv 4\pi\int^r_0\epsilon(r')r'^2 dr'$.
We employ a fourth-order Runge-Kutta method together with a tabulated equation of state obtained from each of the holographic models.

We first consider an EoS calculated in the D4/D8/\textoverline{D8} model.
Physics of dense matter in Sakai-Sugimoto model \cite{Sakai:2004cn} has been developed with/without the source term for baryon charge \cite{denseD4D8,Kim:2007zm}.
One thing that might need to be improved with the Sakai-Sugimoto model for nuclear matter might be the absence of the scalar field which is responsible for
the intermediate attraction of the nuclear force.
In this work we adopt a simple EoS at zero temperature with a point-like baryon source given in \cite{Kim:2007zm}.
Hereafter, we adopt the natural units.
The regularized Helmholtz free energy is given by
\ba
\frac{F_\textrm{reg}(n_B)}{V_3}=a\int_{-\infty}^{\infty} dZ K^{2/3}
\left( \sqrt{1+\frac{(N_c n_B)^2}{4a^2b}K^{-5/3}}-1 \right)\, ,
\ea
where $n_B$ is the baryon number density and
\ba
K=1+Z^2~,~~a=3.76\times10^9 ~{\rm MeV}^4~,~~b=7.16\times10^{-6} ~{\rm MeV}^{-2}\, .
\nonumber
\ea
Here they took $\lambda\simeq 16.71$ and $M_{\rm KK}\simeq 950$ MeV~\cite{Kim:2007zm}.
We adopt the EoS in \cite{Kim:2007zm} to solve the TOV equation.
In this case, however, we could not find a stable compact star,
{\it i.e.}, a star satisfying pressure-zero condition with a radius $R$, $p(R)=0$,
within a reasonable value of the radius. This means that the EoS from the D4/D8/\textoverline {D8} model
may not support any stable compact stars or may support one whose radius is very large.
This might be due to a deficit of attractive force in the D4/D8/\textoverline {D8} since the gravity alone may not balance against the
pressure of the D4/D8/\textoverline {D8} EoS.
In nuclear physics, the long range attractive force is governed by one-pion exchange, while the intermediate
range $\sim 1$ fm attraction is dominated by a scalar field or two-pion exchange.
In the D4/D8/\textoverline {D8}~\cite{Sakai:2004cn} a scalar that couples to baryon fields is missing.
We may find a stable compact star if we try some other EoS based on Sakai-Sugimoto model other than the one given in \cite{Kim:2007zm}.
However, the radius of the resulting
star will be much larger than the one in nature and also the one obtained in the D4/D6 models given below.

\begin{figure}
\subfigure[]{\includegraphics[width=.5\textwidth]{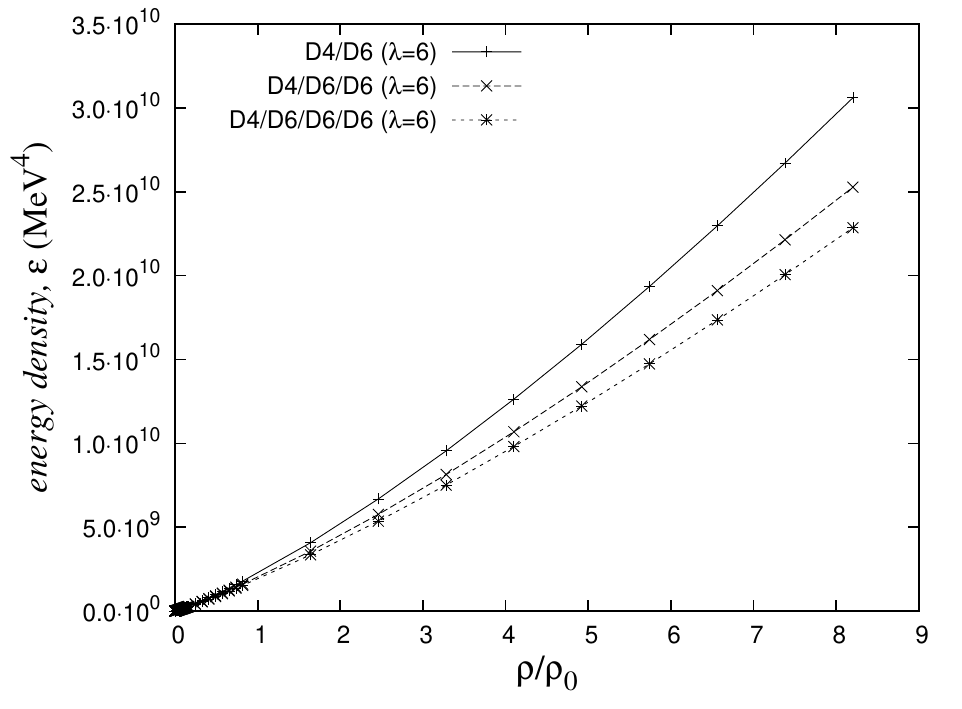}}
\subfigure[]{\includegraphics[width=.5\textwidth]{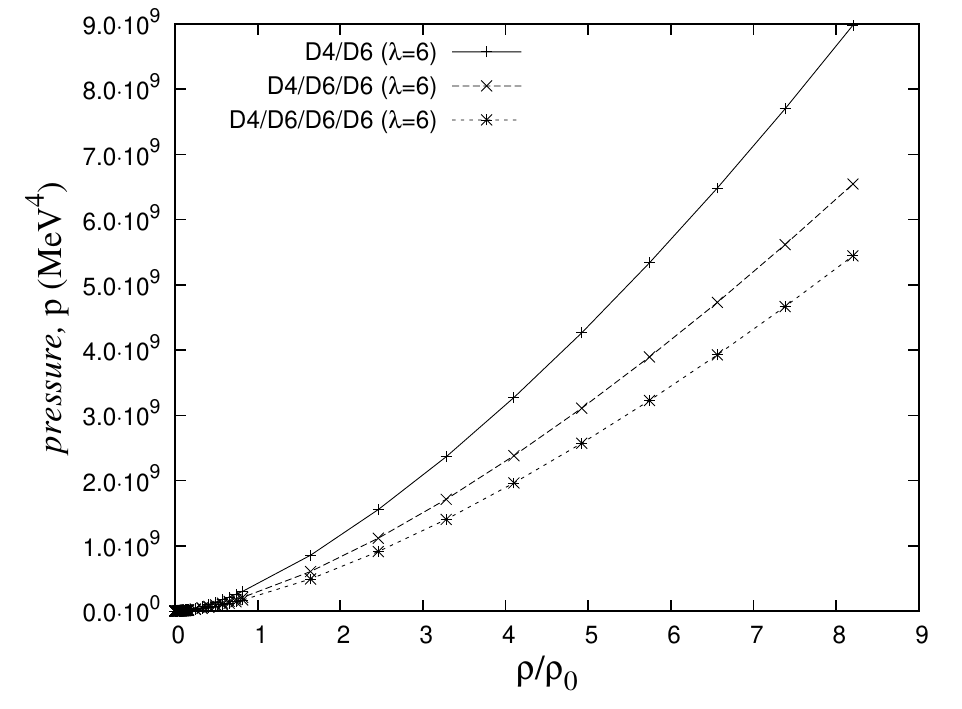}}
\caption{(a): Relation between energy density and baryon number density, and (b): relation between pressure and baryon number density for the three holographic equations of state considered. Here we take $\lambda=6$ and $M_\textrm{KK}=1.04$ GeV.}
\label{fig:EPvRho6}
\end{figure}

Now, we move on to the D4/D6 models.
The EoS is given in Eq. \eqref{EandPinD4D6} as
\begin{eqnarray}
	 \epsilon(\rho)&\!=\!&a\Big(\frac{\tilde{Q}}{4\tau_4}\,\bar{\Omega}_\textrm{D4}+\frac{N_f}{\tau_6}\,\bar{\Omega}_\textrm{D6}^\textrm{reg.}[\tilde{q}]\Big)\nonumber \\
	p(\rho)&\!=\!&a\Big(\frac{N_f}{\tau_6}\,\Omega_\textrm{D6}^\textrm{reg.}[\tilde{q}]\Big)
\end{eqnarray}
where $\rho=b\,\tilde{Q}$ is the baryon number density and
\begin{equation}
	a\simeq4.45\times10^8~{\rm MeV}^4~,~~b\simeq1.07\times10^6~{\rm MeV}^3\,.
\end{equation}
Here we take $\lambda=6$ and $M_\textrm{KK}=1.04$ GeV in D4/D6 model \cite{Jo:2011xq, Jo:2011em}.
Since the value of $M_\textrm{KK}$ is fixed uniquely by a meson mass, $M_\textrm{KK}\sim 1$ GeV is
a rather robust compared to the value of the 't Hooft coupling.
With this choice of our model parameters, we plot the energy density and pressure with respect to the baryon number density
in Fig. \ref{fig:EPvRho6}.
For a given central baryon number density, a configuration is obtained by integrating Eq. \eqref{eq:tov} from the center, $r=0$, to the point where the pressure vanishes, i.e., $p(r=R)=0$, with $R$ being the radius of the configuration.
The mass of this configuration is given by $m(r=R)=M$.
For low baryon number densities, i.e., at $\rho<3\times10^{-4}\cdot\rho_0$, the Baym-Pethick-Sutherland equation of state~\cite{Baym:1971pw}
 is used.
Here $\rho_0$ is the normal nuclear density, and we use $\rho_0= 1.3\times 10^6$ MeV$^3$.
This equation of state is patched smoothly to the holographic equation of state at $\rho \approx 3\times10^{-4}\cdot\rho_0\,{\rm MeV}^3$.

\begin{figure}
\begin{center}
\includegraphics[scale=0.8]{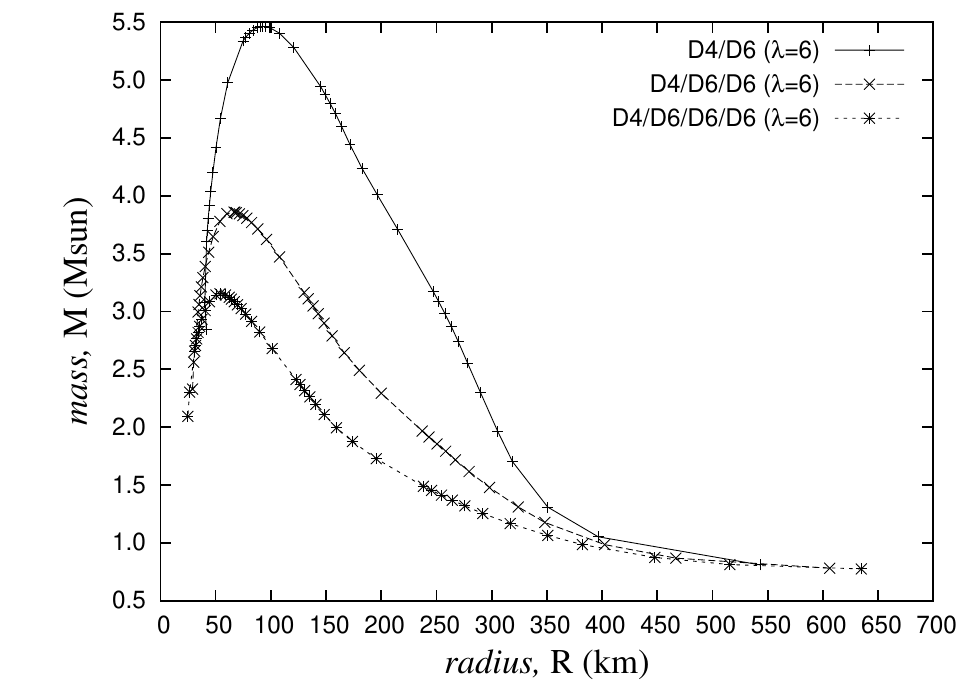}
\caption{Comparison of mass-radius relations for three holographic equations of state with $\lambda=6$ and $M_\textrm{KK}=1.04$ GeV.}
\label{fig:compEOS6}
\end{center}
\end{figure}

By solving the TOV equation, we obtain a mass-radius relation for each equation of state.
In Fig.~\ref{fig:compEOS6}, we compare the mass-radius relations for three holographic equations of state with $N_f=1,2,3$.
As we increase the number of quark flavors we find that the maximum mass and corresponding radius becomes smaller. This
is consistent with the simple picture that larger $N_f$ favors a soft EoS, thereby smaller  maximum mass and  radius.
Roughly speaking, the observed neutron stars have masses around $1.5 M_\odot$ and their radii are around $10$ km.
Compared to these observed neutron stars, what we obtained are more massive and larger.

\begin{figure}
\subfigure[]{\includegraphics[width=.5\textwidth]{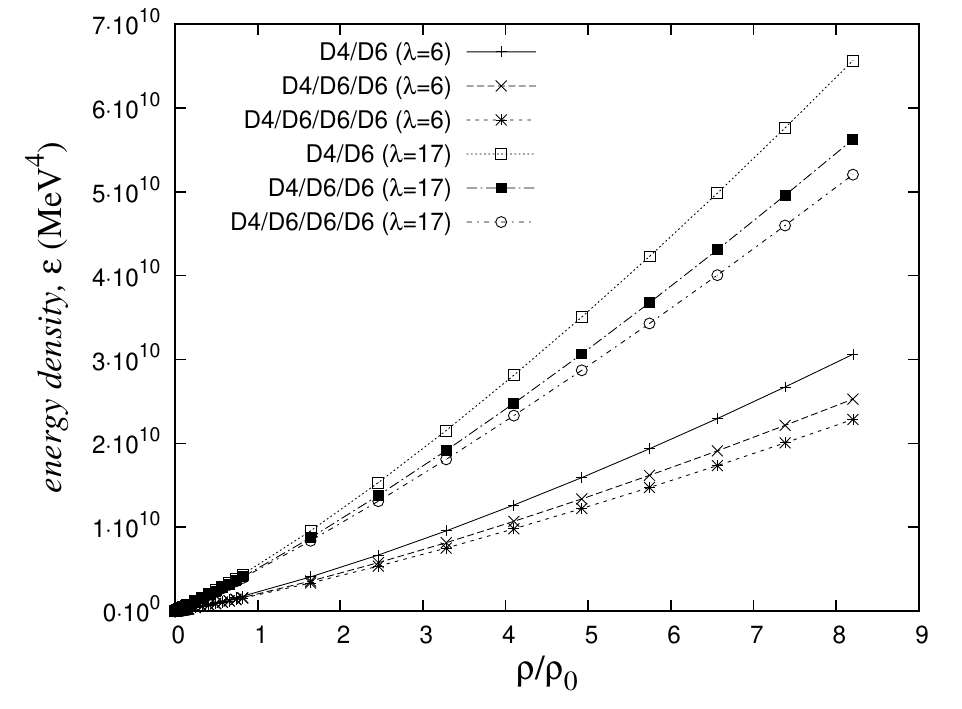}}
\subfigure[]{\includegraphics[width=.5\textwidth]{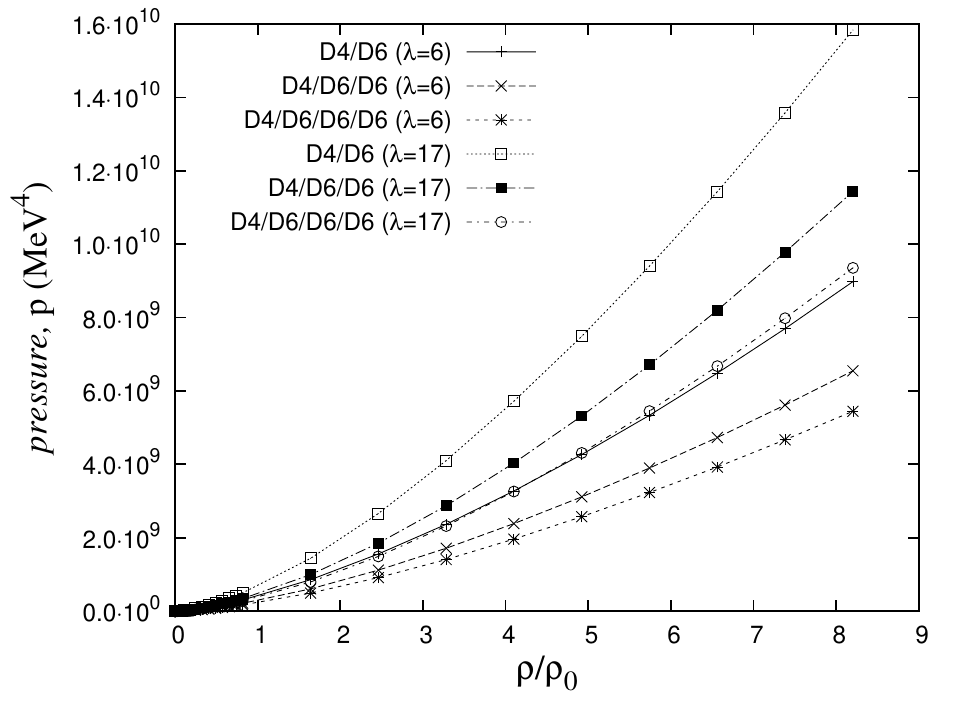}}
\caption{(a): Relation between energy density and baryon number density, and (b): relation between pressure and baryon number density for the three holographic equations of state considered.}
\label{fig:EPvRho}
\end{figure}

\begin{figure}
\begin{center}
\includegraphics[scale=0.8]{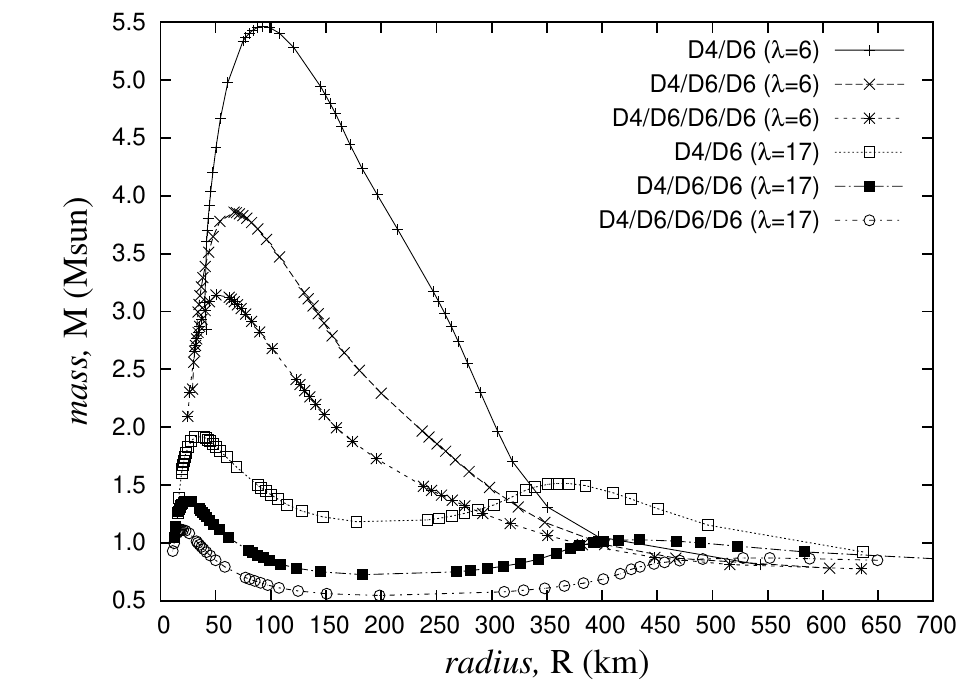}
\caption{Comparison of mass-radius relations for three holographic equations of state.}
\label{fig:compEOS}
\end{center}
\end{figure}

To see the  $\lambda$-dependence of our results, we take $\lambda=17$ and  $M_\textrm{KK}=1.04$ GeV.
We show the energy density and pressure for each model with different values of $\lambda$ in Fig. \ref{fig:EPvRho}.
In Fig. \ref{fig:compEOS}, we collect mass-radius relations with $\lambda=6, 17$ and  $M_\textrm{KK}=1.04$ GeV.
We find that with a bit larger value of the 't Hooft coupling, the mass and radius of our compact stars becomes
smaller. For instance the maximum mass
for $N_f=2$ with $\lambda=6$  was about $3.86 M_\odot$ and its radius was $45$ km, which are now
$M=1.36 M_\odot$ and  $R=17$ km.

\begin{figure}
\begin{center}
\includegraphics[scale=0.8]{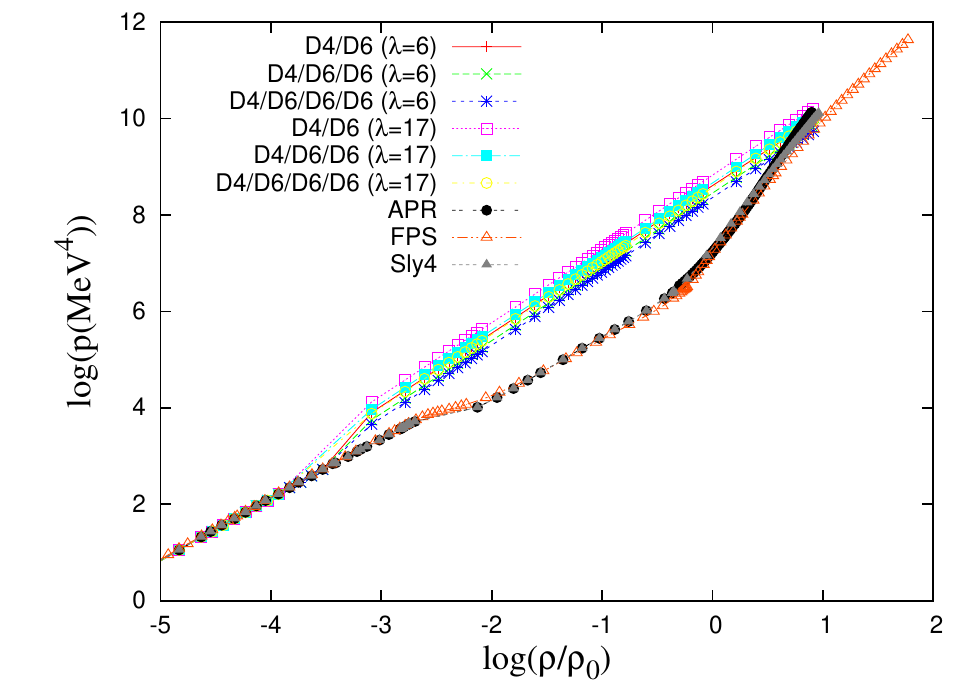}
\caption{Comparison of the pressure from D4/D6 models with those from a few typical EoSs.}
\label{fig:compEOSn}
\end{center}
\end{figure}

\begin{figure}
\subfigure[]{\includegraphics[width=.5\textwidth]{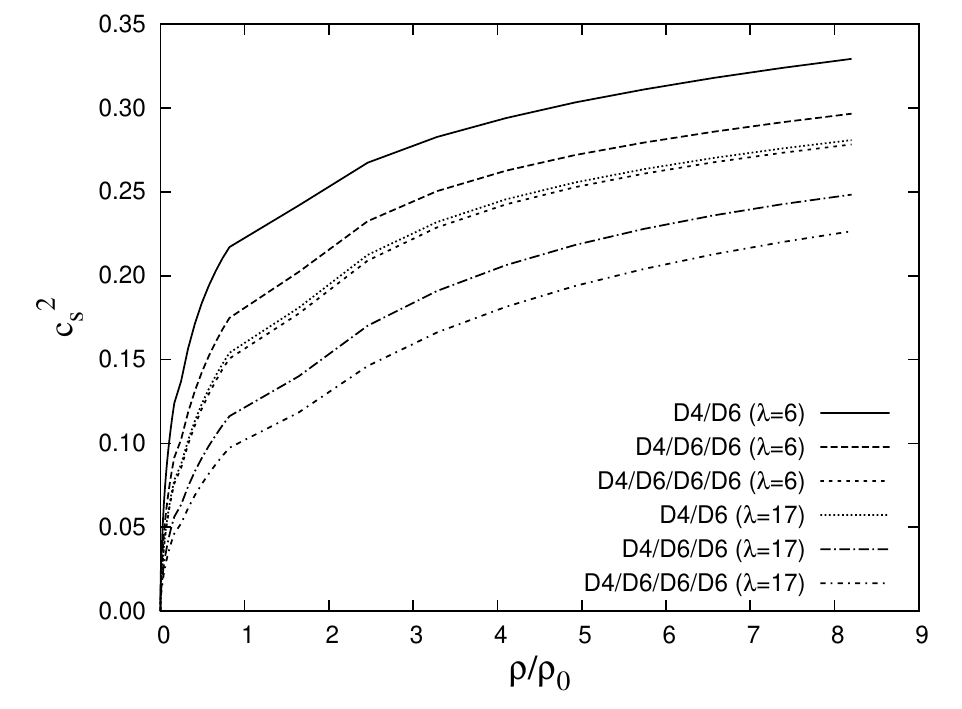}}
\subfigure[]{\includegraphics[width=.5\textwidth]{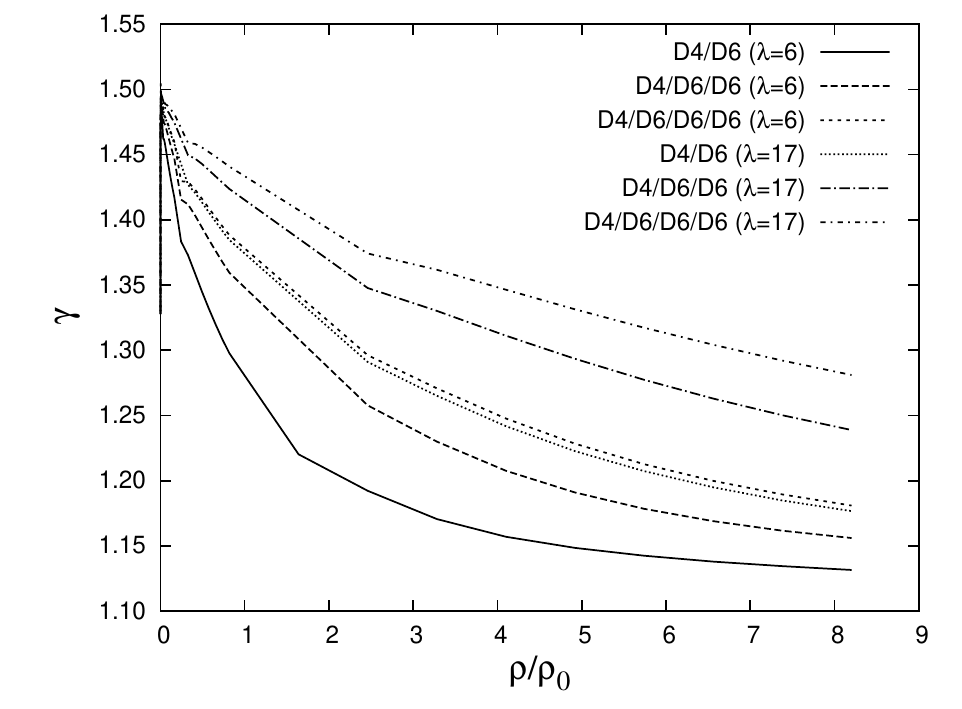}}
\caption{(a) Sound velocity squared $c_s^2$, and (b) the adiabatic polytropic index $\gamma$, for the holographic EoSs with respect to the normalized baryonic number density.} 
\label{fig:sound}
\end{figure}

So far we have focused on the mass-radius relation from holographic EoSs.
Now we look into the holographic EoS in detail to see why our compact stars are different from observed neutron stars and
to check if it satisfies constraints from thermodynamics or causality.
In Fig. \ref{fig:compEOSn}, we compare the pressure from the D4/D6 models with those from a few conventional EoSs: APR~\cite{Akmal:1998cf},
FPS~\cite{FPS}, Sly4~\cite{Sly4}.
Note that pressures from D4/D6 models are quite larger than those from the conventional EoSs in most of the density region that are relevant to neutron stars. This might be the primary reason why our compact stars are quite large in radius compared to other conventional results. 
Large pressure implies large speed of sound. In order to check if the EoSs from D4/D6 models don't violate the causality,
we calculate the square of the sound velocity, $c_s^2=dp/d\epsilon$ on the left panel of Fig.~\ref{fig:sound}.
Our result shows that the speed of sound is less than that for the fully relativistic Fermi gas limit,  $c_s^2=1/3$.
%
%
We also calculated the adiabatic polytropic index $\gamma$ defined by $\gamma=d \ln p/(d\ln \epsilon)$ on the right panel of Fig. \ref{fig:sound}.
Note that the results with $\lambda=17$ shows lower speed of sound but higher adiabatic polytropic index compared to those with $\lambda=6$. This result can be understood from the relation $\gamma=c_s^2 (\epsilon/p)$ and the results in Fig. 3.

%
%

\section{Summary}
We solved the TOV equation using the EoS calculated
in holographic QCD. We first used the EoS from a dense D4/D8/\textoverline {D8} model.
In this case, however, we could not find a stable compact star that satisfies pressure-zero condition $p(R)=0$ at a reasonable value of the radius $R$. This means that the EoS from the D4/D8/\textoverline {D8} model
may not support any stable compact stars or may support one whose radius is very large.
This might be due to the lack of attractive force mediated by a scalar field or two-pion exchange
in the model. Next, we considered the D4/D6 models
with different number of quark flavors, $N_f=1,2,3$.
Though the mass and radius of a holographic star are larger than those of normal neutron stars,
we found stable compact stars in this case. To see why the compact stellar object obtained from the D4/D6 model
is quite different from observed compact stars, we compared the pressure from our model with a few typical ones. We found that the pressure in our work is order of magnitude larger than that of conventional works.
We also confirmed that the EoSs considered in our work don't violate causality by considering the sound velocity square $c_s^2$ and the adiabatic index $\gamma$.
Since holographic QCD is based on the QCD symmetries which are believed to be restored at very high densities, extending the model to low density region may not be valid. In order to have a more realistic compact star, one may have to find a way of introducing proper attraction in holographic QCD or have to make a hybrid star in which low density EoS is replaced by the conventional one.

\acknowledgments
Y.K. thanks Keun-Young Kim and Hyun Kyu Lee for useful discussions.
M.B. Wan thanks Gordon Baym, Feryal Ozel and Christopher Pethick for useful discussions.
Y.K. and I.J.Shin acknowledge the Max Planck Society(MPG), the Korea Ministry of Education, Science and
Technology(MEST), Gyeongsangbuk-Do and Pohang City for the support of the Independent Junior
Research Group at APCTP.
C.H.L. was supported by the BAERI Nuclear R \& D program (M20808740002) of MEST/KOSEF and
the Mid-career Researcher Program through NRG grant funded by the MEST (No. 2009-0083826).

\end{document}